# Charge-driven transtive devices via electric field control of magnetism in a helimagnet

Y. S. Chai[1,2,3], D. S. Shang[2], S. H. Chun[1], Y. Sun[2], Kee Hoon Kim[1]

[1]*CeNSCMR and Institute of Applied Physics, Department of Physics and Astronomy, Seoul National University, Seoul 151-747, South Korea*

[2]*Beijing National Laboratory for Condensed Matter Physics, Institute of Physics, Chinese Academy of Sciences, Beijing 100190, China*

[3]*Low Temperature Physics Laboratory, College of Physics, and Center of Quantum Materials and Devices, Chongqing University, Chongqing 401331, China*

Corresponding author: khkim@phya.snu.ac.kr

**Abstract:**

Transtor and memtranstor are the fourth basic linear and memory elements, which allows direct coupling of charge ($q$) to magnetic flux ($\varphi$) via linear and non-linear ME effects, respectively. It is found here that large variation of magnetization by electric field is realized in both linear and nonlinear hysteretic styles in a magnetoelectric Y-type hexaferrite $Ba_{0.5}Sr_{1.5}Zn_2(Fe_{0.92}Al_{0.08})_{12}O_{22}$ single-crystal. Moreover, based on the spin current model, the underlying microscopic mechanisms for generating the two types of linear and nonlinear $M$ vs $E$ curves are understood as $E$ induced changes of cone angle and sign of $P$ respectively, establishing the charge-driven transtor and memtranstor in the Y-type hexaferrite system. This work points to a promising pathway to develop unique circuit functionalities using the magnetoelectric materials.

# I. INTRODUCTION

In the conventional circuit theory, four basic electrical circuit elements, namely capacitor ($C$), resistor ($R$), inductor ($L$), and memristor can be defined from the relationship among the four fundamental variables: voltage ($v$), current ($i$), charge ($q$) and magnetic flux ($\varphi$) [1-3]. In particular, the memristor is expected to show a nonlinear $i$-$v$ pinched loop. Inspired by the discovery of memristor, memcapacitor and meminductor, the nonlinear memory counterparts of capacitor and inductor accordingly [4], have also been proposed based on the observation of pinched $q$-$v$ and $i$-$\varphi$ loops, respectively. Until very recently, contrary to the resistive switching effects based on the memristor [2], a totally different fourth linear circuit element—transtor ($T$) [5] is proposed to show a direct linear relationship between $q$ and $\varphi$ (Fig. 1(a)) [5]. Its working mechanism is the so-called magnetoelectric (ME) effect, which refers to the induction of electric polarization ($P$) by application of magnetic field ($H$) (direct ME effect) or conversely the induction of magnetization ($M$) by application of electric field ($E$) (converse ME effect) [6-8]. Similarly, memtranstor is also expected to exist with the direct non-linear relationship between $q$ and $\varphi$. With these transtive elements, a unified and complete picture on the relational diagram in the fundamental circuit elements can be established. In other words, a complete circuit diagram is expected to demonstrate both four linear circuit elements, $R$, $C$, $L$, and $T$ and four nonlinear elements, memristor ($M_R$), memcapacitor($M_C$), meminductor ($M_L$), and memtranstor ($M_T$). (See Fig. 1(a) for a pictorial representation) [5].

The ME-based transtive device can be then composed of a block of ME materials sandwiched between two parallel electrodes to operate in either a longitudinal ($E//M$ or $H//P$) or a transverse ($E \perp M$ or $H \perp P$) configuration [5], as shown in the inset of Fig. 2(b). Such a device in the transverse configuration can be proved theoretically to exhibit a direct link between $\varphi$ passing through the flank side and $q$ induced on the electrodes with the following formula:

$$q = DS = (\varepsilon_0 E + P)S = \varepsilon_0 \varepsilon_r ES + \alpha HS \qquad (1)$$

where $D = \varepsilon_0 E + P$ is the electric displacement and $\alpha$ is the ME coefficient. The magnetic flux $\varphi$ passing through the flank side would be:

$$\varphi = BS' = \mu_0(H + M)S' = \alpha ES' + \mu_0 \mu_r HS' \qquad (2)$$

where $B = \mu_0(H + M)$ represents the magnetic induction. Here, $S$ and $S'$ are the area of the ME media beneath the electrodes and that of the flank side, respectively. $\varepsilon_0$ the vacuum permittivity, $\varepsilon_r$ the relative permittivity of the medium, $\mu_0$ is the permeability of vacuum, $\mu_r$ the relative permeability of the medium. From the above equations, we can deduce that:

$$d\varphi = \frac{S'}{S}\frac{\alpha_C}{\varepsilon_0\varepsilon_r}dq \qquad (3)$$

for a charge-driven transtive device based on the converse ME effect, or

$$dq = \frac{S}{S'}\frac{\alpha_D}{\mu_0\mu_r}d\varphi \qquad (4)$$

for a flux-driven transtive device based on the direct ME effect. $\alpha_D = dP/dH$ and $\alpha_C = \mu_0 dM/dE$ are the direct and converse ME coefficient, respectively (they usually have similar values with the same sign). In both cases, the quantity transtance $T = d\varphi/dq$ can be defined to characterize the performance of the ME transtive device. Note that $T$ can be either positive or negative depending on the sign of $\alpha_C$. (Fig. 1(b) for converse ME effect). On the other hand, memtranstor requires a butterfly-shaped hysteresis in the $\varphi$-$q$ loop that corresponds to the hysteresis in the $M$-$E$ or $P$-$H$ loop (only $q$-driven case is shown in Fig. 1(c)).

With those linear and non-linear ME effects, transtive devices have already shown great application potential as a non-volatile multilevel random-access memory [9-12], complete Boolean logic gates [13], synaptic plasticity based neural network model [14-15]. In particular, the charge-driven transtive element employing the converse ME effect would be more practical for the electric circuits. For example, a non-volatile memory based on the non-linear converse ME effect termed as transtance change random-access memory (TCRAM) is proposed [16]. The electric writing of the TCRAM requires a butterfly-shaped hysteresis $M$-$E$ loop in the ME medium to tune the sign and value of $\alpha_C$ and $T$ electrically (Fig. 1(c)). Realization of charge-driven memtranstor would open a pathway for realizing next-generation intelligent devices with circuit functionality. Although the charge-driven device has been realized only in the ME composite structures [9-12], to miniaturize it further, the single-phase ME material would be more favorable. The control of $M$ by $E$ is a difficult task in single phase materials while the realization of a butterfly-shaped hysteresis of $M$-$E$ loop (Fig. 1(c)) is even more challenging. Therefore, the demonstration of charge-driven memtranstor in single phase materials is yet to be realized.

The Y-type hexaferrite BSZFAO consists of a series of Fe/Zn-O tetrahedron and Fe/Al-O octahedron layers stacked along the [001] direction [17]. Under in-plane magnetic field cooling or high magnetic field history, this compound exhibits a commensurate transverse conical ordering based on the previous neutron diffraction study [18], as shown in Fig. 2(a), in which the alternating small ($S$) and large ($L$) magnetic blocks have antiparallel in-plane components with a modulation vector $k_0 = (0, 0, 3/2)$. From the spin-current (or inverse Dzyaloshinskii-Moriya) mechanism [19-20], a ferroelectric polarization can be induced by the transverse cone, where $P \propto \sum_{ij} k_0 \times (S_i \times S_j)$, and $S_i$ and $S_j$ are spins on the two adjacent sites along [001] direction. The induced $P$ will be always perpendicular to $H$ and $k_0$, as shown in Fig. 2(a). In the transverse cone phase of BSZFAO, a giant ME coefficient $\alpha_D \sim 2 \times 10^4$ pm/s has been found at zero-$H$ at 30 K [21]. Conversely, an $E$ control of large $M$ reversal at zero-$H$ has also been realized in BSZFAO below 170 K due to its large $\alpha_C$ [22]. Therefore, it is a good candidate material to realize the functionalities of charge-driven transtor and memtransor.

In this work, we have investigated $E$ control of $M$ under large biased−DC magnetic field in an ME Y-type hexaferrite Ba$_{0.5}$Sr$_{1.5}$Zn$_2$(Fe$_{0.92}$Al$_{0.08}$)$_{12}$O$_{22}$ (BSZFAO) single crystal. Varying the strength of $H$ bias, both linear and non-linear converse ME effects are selectively controlled. Our finding uncovers that BSZFAO can serve as a ME medium for the charge-driven transtor and memtransor based on the linear and the butterfly-shaped $\varphi$-$q$ relationships, respectively.

## II. Methods

The Y-type hexaferrite Ba$_{0.5}$Sr$_{1.5}$Zn$_2$(Fe$_{0.92}$Al$_{0.08}$)$_{12}$O$_{22}$ single crystals were grown from a Na$_2$O-Fe$_2$O$_3$ flux in air and annealed at 900ºC in flowing oxygen to enhance the resistivity [21]. The starting chemicals were mixed with a molar ratio of BaCO$_3$: SrCO$_3$: ZnO: Fe$_2$O$_3$: Al$_2$O$_3$: Na$_2$O=2.95: 16.74: 19.69: 49.32: 4.29: 7.01. A single crystal was cut into a rectangular shape for electrical measurements along [120] direction while $H$ is along [100] direction so that $E \perp H \perp k_0$, as shown in the inset of Fig. 2(b). Towards this end, the temperature and magnetic field were supplied by a Physical Property Measurement System (Quantum Design, USA). In a magnetoelectric poling procedure, the magnetic field was applied along the [100] direction, and the electric field was applied along the [120] direction of the sample. In order to achieve a magnetoelectric poling condition of either +Poled or –Poled status, the electric field of $E$= 220 kVm$^{-1}$ was applied to the sample at 120 K, while the magnetic field was varied from 20 kOe to 12 kOe to bring the crystal from paraelectric/collinear phase to the ferroelectric/transverse cone phase. Then the sample was cooled down to 15 K. Finally, the electric field was first turned off, and then the magnetic field was ramped to the demanded value for the forthcoming electric measurements. The magnetoelectric current was measured and integrated to determine the change in electric polarization with the magnetic field by a high resistance electrometer (Keithley 6517B). The same procedure was adopted for the measurement under –Poled condition, with the direction of electric field reversed. Magnetization curves were measured using a vibrating sample magnetometer (VSM) in a PPMS (Quantum Design). For the magnetization measurements under a bias electric field, the sample temperature would be maintained at 15 K as for poling, while the magnetic field would be varied from 12 kOe, as for poling, to a demanded value. For the magnetization measurements under electric field, the sample holder of the conventional vibrating sample magnetometer was modified to tolerate a high electric field.

### III. RESULTS AND DISCUSSION

#### A. Magnetic and direct ME properties of BSZFAO

To reveal the ferroelectricity of BSZFAO, the target temperature is chosen to be 15 K where the BSZFAO is reported to show $H$-induced ferroelectricity and strong ME coupling below 20 kOe [21]. At 15 K, $H$-dependent electric and magnetic measurements are performed, as shown in Fig. 2(b) and 2(e). In Fig. 2(b), the $H$-dependent relative dielectric constant $\varepsilon_r$ exhibits two clear peaks at around 0 and 19 kOe respectively, where a transverse cone and ferroelectric phase is expected to exist in between those fields. Accordingly, the $H$-dependent $M$ curve also shows two kinks at similar $H$ values, as seen in Fig. 2(c). As $H$ increases, the $M$-$H$ curve shows a rapid increase from 0 to ~0.5 kOe, less sharply increases from 0.5 kOe to 19 kOe and then almost saturates after 20 kOe. Above 19 kOe, BSZFAO enters the ferrimagnetic phase [18]. The existence of ferroelectricity between 0 and 19 kOe is confirmed with the non-zero $P$ by integrating the ME currents as a function of time (Fig. 2(d)) as well as the sign reversal of $P$ by the reversed poling of $E$ in this field region. The sign of $P$ in a transverse cone state is then determined by the sign of spin helicity $\sum_{ij} S_i \times S_j$; namely, the sign is determined by whether the spins in the transverse cone phase rotate in either clockwise or counter clockwise directions as shown in the inset of Fig. 2(c). Above 19 kOe, ferrimagnetic phase is paraelectric (PE) because $S_i \times S_j = 0$ for collinear spins.

Moreover, there are strong *H* dependence in the *P* behaviors in the FE phase. The absolute *P* value (|*P*|) shows a rapid increase below 0.5 kOe and reaches a plateau between 0.5-3 kOe (Fig. 2(d)). However, further increase of *H* decreases |*P*| gradually and finally makes it zero above 19 kOe where BSZFAO is paraelectric. Accordingly, huge non-zero d*P*/d*H* (=$\alpha_D$) can be calculated, as shown in Fig. 2(e). For +*P* case, the $\alpha_D$ shows a giant positive peak centred at zero *H* (not fully shown here), quickly approaches to zero above 0.5 kOe, monotonically decreases and becomes negative between 0.5 kOe to 18 kOe before vanishing above 19 kOe. The large $\alpha_D$ within the FE phase should have magnetic origins. The zero-field peak in $\alpha_D$ comes from the in-plane rotation of the transverse cone domains between ±3 kOe [22]. However, the sign reversal of $\alpha_D$ value between 3 and 19 kOe must have different physical origin. One can easily expect that with increasing *H*, the transverse cone open angle $\theta$ will decrease accordingly (inset of Fig. 2(d)), thus decreasing the *P* values induced by the spin-current mechanism. It can naturally explain the simultaneously increase of *M* and decrease of *P* in the field range of 3-19 kOe. Finally, the peaks near 18 kOe in d*P*/d*H* curves indicate the onset *H* of the transition from the transverse cone to the collinear ferrimagnetic phases.

Following Eq. (2), the non-zero $\alpha_D$ observed in BSZFAO means that a flux-driven transtor, with a device configuration in the inset of Fig. 2(b) and using the crystal as the working medium, can be actualized and operated in the parameter domain specified above. The different signs of $\alpha_D$ for +Poled and -Poled cases result in different sign of transtance *T*. As we mentioned above, it is more challenging to realize the charge-driven transtive devices since the demonstration of *E* control of *M* is much harder and rarer than that of *H* control of *P*. However, we have already demonstrated in BSZFAO a giant d*M*/d*E*, *i.e.* converse ME effect at zero-*H* [22]. The converse ME effect at zero-*H* shows small hysteresis (Fig. 3(a)), making it not ideal for either transtor or memtranstor [5]. By applying small dc bias magnetic field, the hysteresis behavior of *M-E* loops is suppressed while its linear ME behavior is replaced by a quadratic ME behavior (Fig. 3(a)), which is also unfavourable to the transtive devices. In the following, we will show that by applying stronger dc *H* bias, we could still realize nearly ideal charge-driven transtor and memtranstor devices based on BSZFAO.

### B. Linear converse ME effect and charge-driven transtor of BSZFAO

We first applied a constant bias $H$ = 10 kOe to drive the system to the single magnetic transverse cone domain state having a finite $\alpha_D$. A slowly varying $E$ is found to induce a nearly linear $M$ response, yielding a non-zero $\alpha_C$ (=$\mu_0 dM/dE$ in this case), as shown in Fig. 3(b). $\alpha_C$ also allows both positive and negative values, depending on the poling condition while for the same poling condition, $\alpha_C$ and $\alpha_D$ share the same sign. No hysteresis or very small quadratic component is shown in such curves. To understand the underlying physics, the effects of $E$ on $M$ can be considered as an inverse process of $H$ over $P$. As schematically illustrated in Fig. 3(d), the electric field can change the magnitude of the polarization and indirectly causes a change in the transverse cone open angle $\theta$ or in other words, a change in magnetization. Depending on the different initial ME poling condition, the $E$ can enhance or reduce $M$, resulting in opposite sign of $\alpha_C$, which is consistent with our observations.

With the help of the above scenario, the $M$-$E$ curves in BSZFAO can be readily converted to $\Delta\varphi$-$\Delta q$ response curves in the device made of BSZFAO, as shown in Fig. 3(c). For the data in Fig. 3(b) with dc $H$ = 10 kOe, we can easily obtain that $\Delta q_{ME} = \varepsilon_0\varepsilon_r\Delta ES$ from Eq. (1). From Eq. (2), $\Delta\varphi = \mu_0(\Delta H + \Delta M)S' = \mu_0\Delta MS'$. Its transtance ($T=\Delta\varphi/\Delta q$) is either positive or negative depending on the poling condition. The above $\Delta\varphi$-$\Delta q$ relationship perfectly fits to a typical charge-driven transtor. Therefore, such a mechanism makes other ME hexaferrites with a transverse cone state good candidates for the charge-driven transtors.

### C. Nonlinear converse ME effect and charge-driven memtranstor of BSZFAO

Additionally, the same device turns out to be a charge-driven memtranstor when the bias dc $H$ is set to be 18 kOe where the |d$P$/d$H$| shows a peak feature. This $H$ value can drive the BSZFAO crystal approaching the FE-PE phase boundary. In this case, the magnetic response became double-periodically tuned to the driving $E$ after the first 12.5 seconds, and it was always nonlinear (Fig. 4(a)). By redrawing the time-varying data of $E$ and $M$ in Fig. 4(a) to the $M$-$E$ curves in Fig. 4(b), a butterfly-shaped loop can be clearly identified, with the crossing point of the curves for up- and down-sweeps of $E$ in the vicinity of $E$ = 0. Such a butterfly-shaped $M$-$E$ loop under constant $H$ corresponds to a butterfly-shaped $\Delta\varphi$-$\Delta q$ loop in Fig. 4(d) for the model device, where $\Delta q = (\varepsilon_0\Delta E + P)S = \varepsilon_0\varepsilon_r\Delta ES$ and $\Delta\varphi = \mu_0(\Delta H + \Delta M)S' = \mu_0\Delta MS'$. The opposite slope for the $\Delta\varphi$-$\Delta q$ curves in the region from point **1** to point **2** and that from point **3** to point **4** are referred to the opposite signs of transtance *T* values. Therefore, our device with the BSZFAO crystal indeed demonstrates a characteristic behavior of a charge-driven memtranstor.

To understand further the physical process involved in the butterfly-shaped *M-E* loop, we carefully check the data in Fig. 4(b). Within point **1** to point **2**, the $\alpha_C$ <0 indicates a positive *P* in this region, as illustrated in the upper panel of Fig. 3(d). From point **2** to point **3**, $\alpha_C$ suddenly becomes positive, indicating a negative *P*, *i.e.*, the polarization has been reversed by negative electric field. In the region between point **3** and point **4**, $\alpha_C$ >0 again, confirming that the negative *P* was maintained during the process. Finally, from point **4** to point **5**, $\alpha_C$ suddenly becomes negative, indicating a positive *P*, *i.e.*, the polarization has been reversed again by positive electric field. The above processes are schematically demonstrated in the Fig. 4(c). In the meantime, *P* reversal implies the reverse of spin helicity due to the spin current mechanism. In other multiferroics with spiral or conical ordering, *e.g.*, TbMnO$_3$ and MnWO$_4$, the reversal of spin helicity upon the reversal of *P* has already been reported in neutron diffraction studies [23,24]. It is very likely to expect a hysteresis of polarization following slow *E*-sweeping near FE-PE phase boundary. Usually, for a ferroelectrics, the coercive field is zero at the phase boundary and it becomes larger when it moves away from the boundary. In the Y-type hexaferrite we studied, the FE phase resides between ±19 kOe at 15 K so that the high magnetic field is closer to the boundary while the low magnetic field is further from the boundary. Therefore, the butterfly curve appears at high magnetic field with low ME coupling, but not at low fields with high ME coupling. In our case, the *E* values at points **2** and **4** represent the negative and positive coercive *E* respectively (Fig. 4(c)).

In this material, the boundary between FE-PE persists up to at least 220 K [21] where the butterfly-shaped *M-E* loop is allowed. However, due to the leakage nature of this sample in high temperature, we could not utilize high enough *E* to observe this hysteresis behaviour. The above underlying physics demonstrates a memtranstive principle, which can be readily applicable to other single phase ME systems with a heliconical spin ordering. There are plenty of magnetoelectric hexaferrites which show heliconical spin configuration under magnetic field and many of these can persist this order around room temperature or even higher with a substantial ME coupling [25]. Moreover, in the domain reversal processes, the relaxation mechanism would play a role since more domain with opposite *P* will be reserved with longer time. The time scale of spin reversal process can be very fast which points to a much short device on-off period.

It is worth pointing out that the memtranstor demonstrated in BSZFAO does not have an ideal butterfly-shaped loop, *i.e.*, not pinched at $E = 0$, due to the existence of tiny spontaneous polarization. From Eq. (1), spontaneous polarization will supply a non-zero initial value of $q$, leading to a crossing of the hysteretic loop slightly shifted from zero *E*. Even with the deviation from a zero-driving field for the hysteresis loops, the behaviors exhibited by our model device using the ME effect, or specifically the differential relations between $\varphi$ and $q$, still meet the definition for the fourth fundamental memelement in an extended sense. Similar behavior has been observed in other memristive devices so that the memristive, memcapacitive, and meminductive systems can be extended into the dynamic cases where pinching occurs elsewhere from the zero-driving field [26].

## IV. CONCLUSIONS

Electric field control of large magnetization change is realized in both linear and butterfly-shaped hysteresis patterns with two underlying mechanisms in a magnetoelectric Y-type hexaferrite $Ba_{0.5}Sr_{1.5}Zn_2(Fe_{0.92}Al_{0.08})_{12}O_{22}$ single-crystal at 15 K. Based on this magnetoelectric material system, the expected fourth circuit elements, charge-driven transtor and memtranstor of linear and non-linear relationship between $\varphi$ and input $q$, respectively, are unambiguously demonstrated via the achieved converse magnetoelectric behaviors. The two mechanisms of *E* control of *M* will be very helpful for developing more novel circuit functionalities with the ME hexaferrites in future.

**Acknowledgments**

This work was supported by the National Key Basic Research Program of China grant No. 2011CB921801, and by the Natural Science Foundation of China grant No. 11974065. The work at SNU was supported by the National Research Foundation (NRF) of S. Korea through 2019R1A2C2090648, 2019M3E4A1080227, and 2021R1A6C101B418.

**Figure captions**

**FIG. 1**. (a). Complete relational diagram of all the possible fundamental two-terminal circuit elements, both linear and nonlinear which correlate a particular pair of the four basic circuit variables, *i.e.*, the charge $q$, the voltage $v$, the current $i$, and the magnetic flux $\varphi$. The nonlinear memory devices corresponding to the four linear fundamental elements, *i.e.*, the resistor ($R$), the capacitor ($C$), the inductor ($L$), and transtor ($T$), are accordingly termed as memristor ($M_R$), memcapacitor ($M_C$), meminductor ($M_L$), and memtranstor ($M_T$). (b). Schematic illustration of the charge driven transtive behaviors which is anticipated to directly correlate $\varphi$ with $q$ with a positive or negative $T$ values. (Or equivalently, a linear change of $M$ under $E$ with either positive or negative $\alpha_C$ in ME materials.) (c). Characteristic behavior of the charge driven memtranstor showing a butterfly hysteresis loop between $q$ and $\varphi$. The butterfly-shaped loop in (c) occurs at a sufficiently large input of $q$. (Or equivalently, a butterfly shape of $M$ under $E$ in ME materials with a large enough $E$ to induce a sign change in $\alpha_C$.)

**FIG. 2** (a). Schematic illustration of the crystal and transverse conical magnetic structure of $Ba_{0.5}Sr_{1.5}Zn_2(Fe_{0.92}Al_{0.08})_{12}O_{22}$. A transverse conical spin configuration propagates along a commensurate modulation vector $k_0 = (0,0,3/2)$, $\mu_L$ and $\mu_S$ denote the net magnetic moments in the magnetic $L$ and $S$ blocks respectively. According to the spin-current mechanism, the ferroelectricity (FE) and an in-plane $P$ perpendicular to both $H$ and $k_0$ can be induced. Magnetic field dependence of (b) relative dielectric constant $\varepsilon_r$, (c) magnetization $M$, (d) polarization $P$ and (e) d$P$/d$H$ of $Ba_{0.5}Sr_{1.5}Zn_2(Fe_{0.92}Al_{0.08})_{12}O_{22}$ at 15 K after magnetic field cooling processes. The measurement configuration is illustrated in the inset of (b). In the inset of (d), the cone open angle $\theta$ is expected to be suppressed with increasing $H$ between 3 and 19 kOe. The sign of $P$ is directly correlated to the spin rotation directions. Above 19 kOe, the spin configuration becomes collinear that it is paraelectrics (PE).

**FIG. 3** (a). Electric field ($E$) dependence of $M$ under different low dc magnetic field ($H$) at 15 K. Clear hysteresis behaviors can be seen in $M$-$E$ loops for $H = 100, 0, -100$ Oe. Strong quadratic ME behaviors are clear for $H = \pm 200$ Oe. The data at $H = 0$ Oe are brought from Ref. [14]. (b). Nearly linear ME behaviors for both +Poled and –Poled cases under $H = 10$ kOe. (c). Converted $\phi$-$q$ relationship (from (b)) of the transtor made from $Ba_{0.5}Sr_{1.5}Zn_2(Fe_{0.92}Al_{0.08})_{12}O_{22}$. (d). Schematic diagrams of the change of cone open angle under external $E$ for +Poled (upper panel) and –Poled (lower panel) case. Solid and dash lines represent the cone, $P$ and $M$ under +$E$ and –$E$ respectively.

**FIG 4.** (a). Non-linear modulations of magnetization $M$ at 18 kOe and 15 K by repeating triangular waves of $E$ after ME +Poled. From point **1** to **5**, the $E$ is swept from 2.1 MV/m to -2.1 MV/m and back to 2.1 MV/m again. (b). Butterfly-shaped $M$-$E$ loop illustrating the data in (a) spanning the range from point **1** to **5**. (c). Schematic diagram of $P$-$E$ loop from point **1** to **5**. Note that point **1** and point **3** should have different cone rotation directions. Accordingly, their signs of $P$ and $\alpha_C$ are opposite. (d). The butterfly-shaped $M$-$E$ loop in (b) can be directly converted to a memtranstive $\varphi$-$q$ loop.

**Figures**

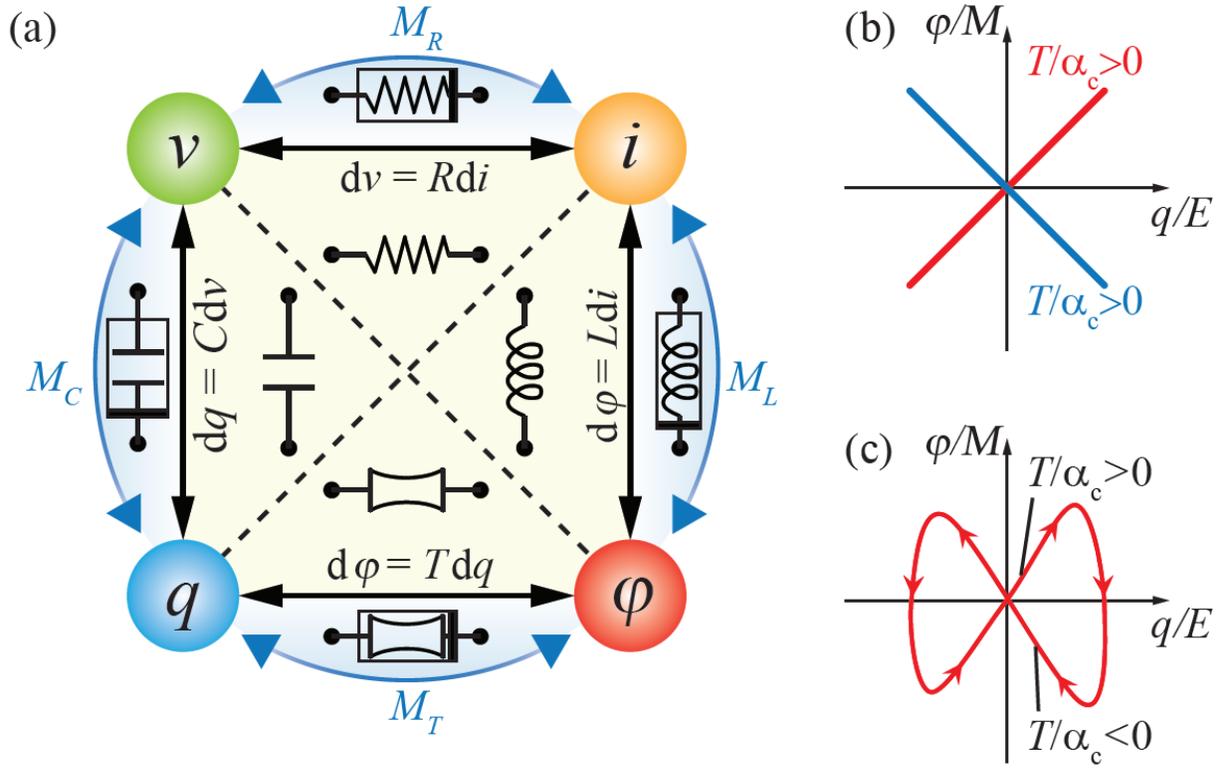

**Figure 1**

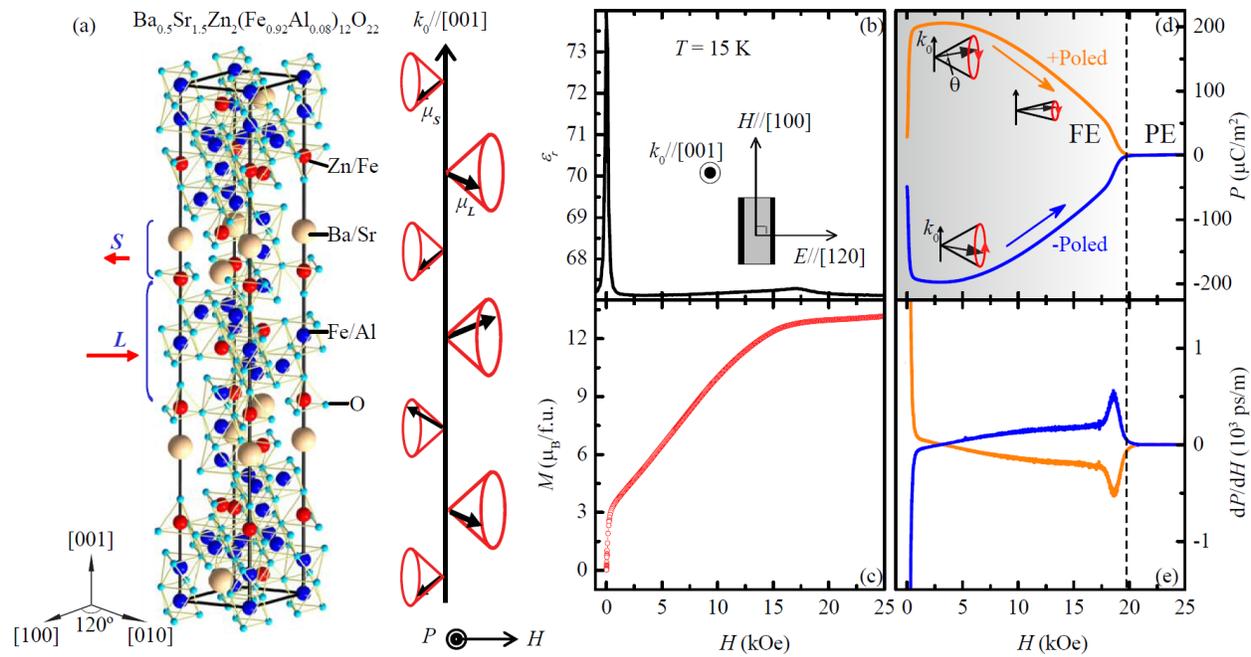

**Figure 2**

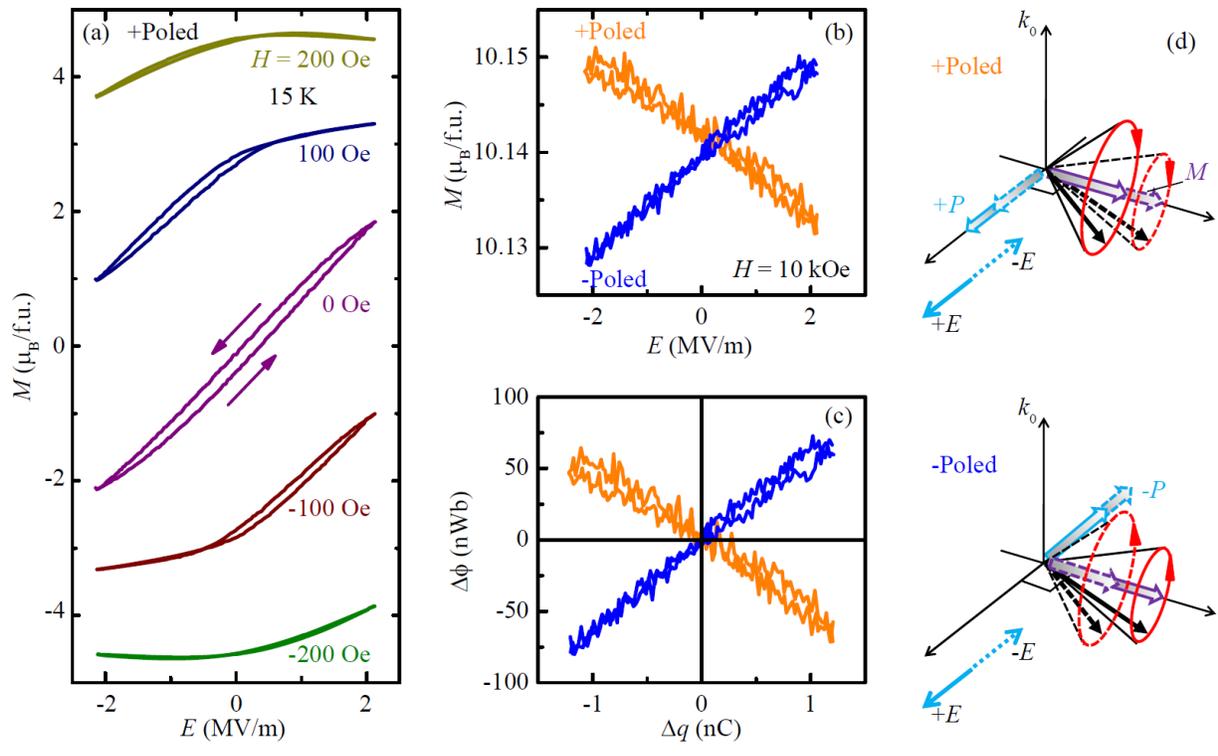

**Figure 3**

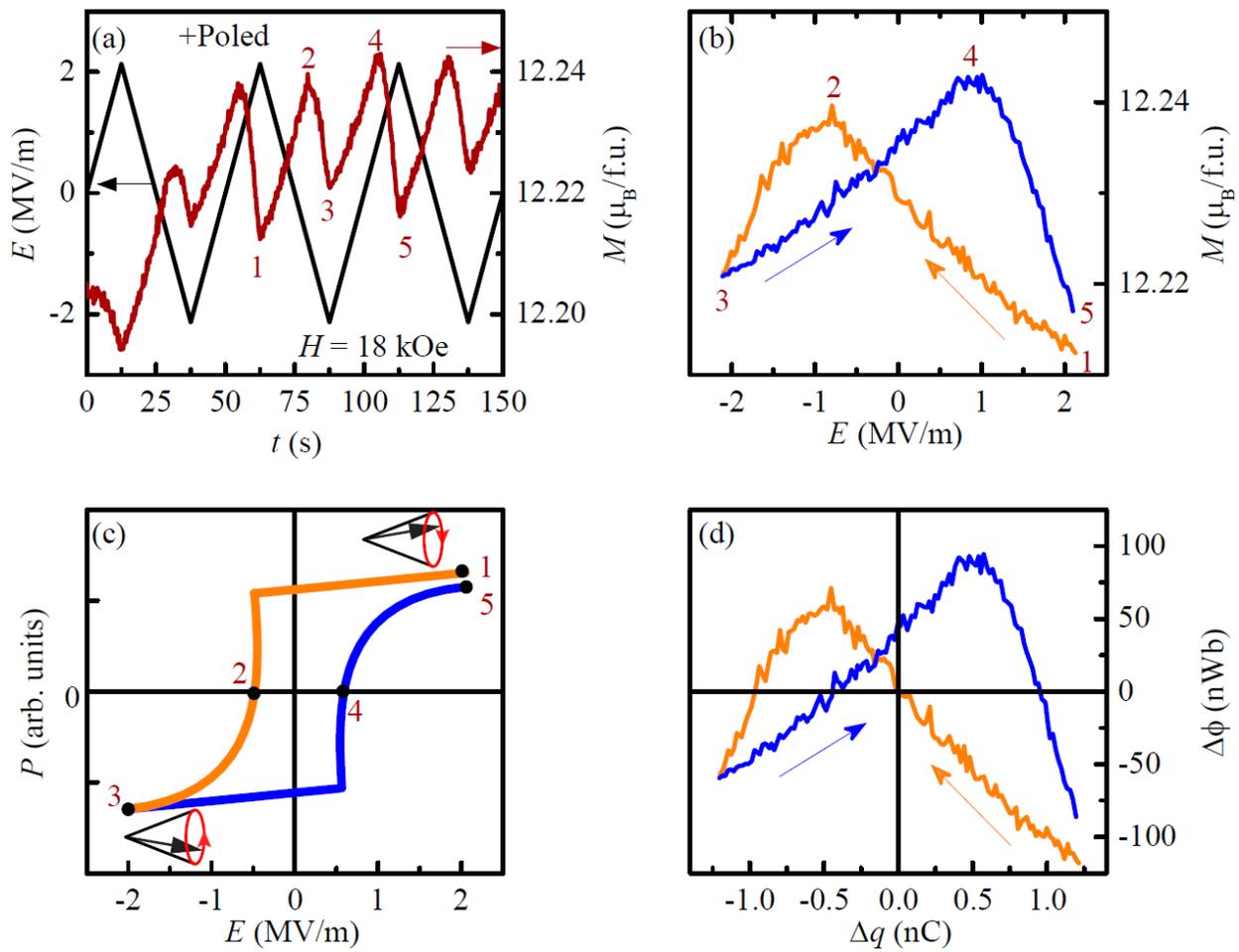

**Figure 4**